\documentclass{ws-ijmpb}

\def\bfu{\mbox{\bf u}}
\def\bfB{\mbox{\bf B}}
\def\bfe{\mbox{\bf e}}
\def\bfr{\mbox{\bf r}}
\def\dd{{\rm d}}
\def\bfnabla{\mbox{\boldmath $\nabla$}}

\newcommand*{\Ray}{{\rm Ra}}
\renewcommand*{\Pr}{{\rm Pr}}
\newcommand*{\Pm}{{\rm Pm}}
\newcommand*{\q}{{\rm q}}
\newcommand*{\Ek}{{\rm E}}

\begin{document}

\markboth{Morin \& Dormy}
{Dynamo bifurcation}

%
\catchline{}{}{}{}{}
%

\title{The dynamo bifurcation in rotating spherical shells}

\author{VINCENT MORIN \& EMMANUEL DORMY}

\address{MAG (CNRS/ENS/IPGP), Ecole Normale Sup\'erieure,\\ 24 rue Lhomond,
  75252 Paris Cedex 05, France\footnote{VM is now at: Laboratoire de Physique, Ecole Normale Sup\'erieure
de Lyon, CNRS UMR5672, 46 all\'ee d'Italie, F-69364 Lyon, France.}\\[2mm]
vincent.morin@ens-lyon.fr, dormy@phys.ens.fr}

\maketitle

\begin{history}
\received{21 Oct. 2009}

\end{history}

\begin{abstract}
We investigate the nature of the dynamo bifurcation in a configuration
applicable to the Earth's liquid outer core, i.e. in a rotating spherical shell with thermally
driven motions.
We show that the nature of the bifurcation, which can be either
supercritical or subcritical or even take the form of isola (or detached
lobes) strongly depends on the parameters. This dependence is described in
a range of parameters numerically accessible (which unfortunately remains remote from
geophysical application), and we show how the magnetic Prandtl number
and the Ekman number control these transitions.
\end{abstract}


\section{Introduction}

The origin of the Earth magnetic field remains a challenging issue for
physicists. The accepted theory, known as dynamo theory, describes the
transfer from kinetic to magnetic energy as an instability process. Above a
given threshold electrical currents, and thus magnetic fields, are amplified
by a turbulent flow of a conducting fluid. 

This problem can be described in its simpler form by a set of coupled
partial differential equations (see section~2). The state of this system is
fully characterised by four independent controlling parameters. The Ekman
number $\Ek$, which can be interpreted as measuring the ratio of the length of
the day to a typical viscous timescale, this number is extremely small in
the case of the Earth's core (the Earth is in rapid rotation). The magnetic
Prandtl number $\Pm$, measuring the ratio of a typical ohmic timescale to 
viscous timescale, it is a characteristic of the fluid and is minute for
all liquid metals (including liquid iron in the Earth's core). The
hydrodynamic Prandtl number $\Pr$, also characterizing the fluid and which
compares a thermal timescale to a viscous timescale, this number is of
order one, or slightly less, but non-vanishing. Finally the Rayleigh number $\Ray$,
which measures a ratio of driving forced to forced slowing down the motion. Its
value is difficult to assess in a simple Boussinesq model. This will be the 
most obvious controlling parameter, which needs to be varied to investigate
dynamo properties.

\begin{figure}
\begin{center}
\includegraphics[width=0.65\linewidth]{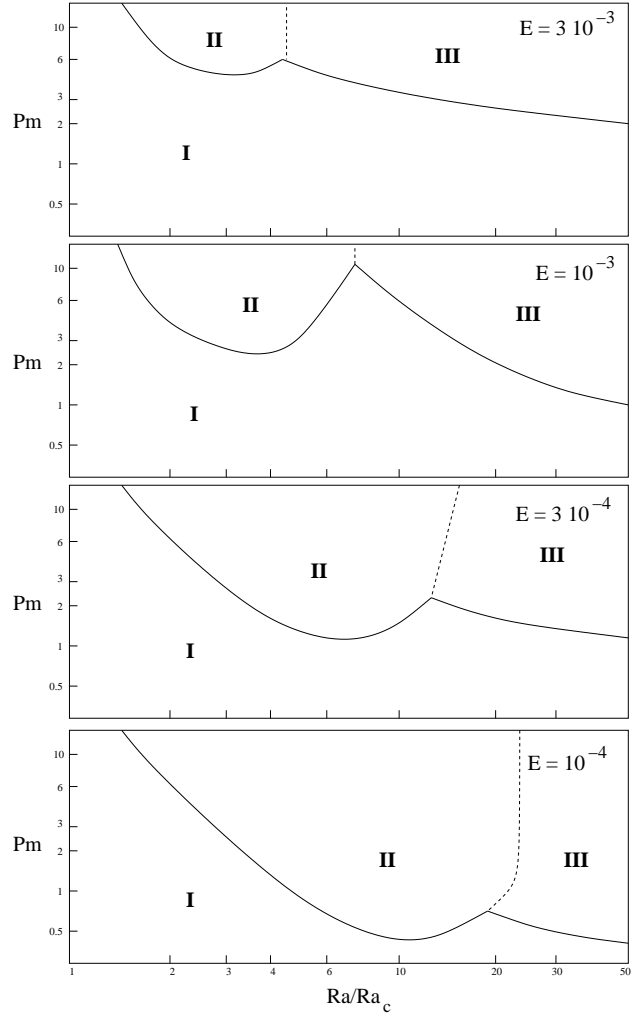}
\caption{``Phase diagram'' for dynamo action for decreasing values of the
  Ekman number (after Kutzner, 2003).
Three domains are identified. Domain {\bf I} corresponds
to parameters for which no dynamo solutions were obtained.
Domain {\bf II} corresponds to dynamo solutions dominated by a strong axial
dipole. Domain {\bf III} corresponds to reversing multipolar dynamos.
}
\label{kutzner}
\end{center}
\end{figure}

Kutzner and Christensen [\refcite{KC02}] performed a first systematic exploration of
the parameter space available to numerical models of the geodynamo.
They produced a ``phase diagram'' for dynamo action in terms
of Ekman number $\Ek$ and magnetic Prandtl number (or equivalently here
Roberts number, $\q=\Pm/\Pr$) and Rayleigh number $\Ray$.
Their study was later extended to lower values of $\Ek$ and $\Pm$ in~[\refcite{Aubert}].
We present a schematic sketch of their results in
figure~\ref{kutzner} (See also the PhD thesis of Carsten 
Kutzner~[\refcite{CK03}]).
Each diagram corresponds to a given Ekman number, ranging here from
$2.10^{-2}$ to $10^{-4}$. 
Kutzner and Christensen identified three ``phases'' on these diagrams. In
part of the parameter space, no dynamo solutions were obtained. In the complementary
region, in which dynamo is obtained, two regimes are identified. One at low
forcing, characterised by strongly dipolar, but non-reversing dynamos, and
another corresponding to reversing, but multipolar dynamos. This, of
course, is not a very good news when comparing these models with the geodynamo
(characterized by a reversing strongly dipolar field).

As pointed above, all models presented in these diagrams suffer from
parameters extremely remote from their geophysically relevant estimates. 
In particular, the Ekman number $\Ek$ is over-estimated by a factor in excess
of a billion and the magnetic Prandtl number $\Pm$ by a factor close to a million. In
fact this last number offers a ratio of diffusivity which is equal to unity 
or larger in the numerics, whereas it should be as small as $10^{-6}$
in the Earth. The diagram reveals that if the magnetic Prandtl number is
decreased in numerical models (all other parameters being fixed), dynamo
action is lost! This does not come as a surprise, as it corresponds to
increasing the ohmic resistivity, but appears to prevent any direct comparison of
the models with the actual geodynamo.

In fact the main result derived from these phase diagrams is much more
optimistic. The authors noted that the threshold value $\Pm_L$, below which
dipolar dynamos are lost decreases with the Ekman number. They proposed~[\refcite{COG,Aubert}] a
scaling in the form $\Pm_L \propto \Ek^{\alpha}$ and proposed a numerical fit
to $\alpha\simeq 3/4$ (see~[\refcite{CRAS}] for a 
theoretical explanation of this behaviour leading to $\alpha\simeq
2/3$). Which could indicate a connection in the parameter space between the 
models presently available and the regime relevant to the geodynamo.

Part of the Kutzner \& Christensen results have also been discussed
in~[\refcite{COG}]. In this earlier study, the authors 
proposed that for $\Ek=10^{-3}$ the  ${\bf B}={\bf 0}$ state is always
stable.  The corresponding bifurcations are therefore interpreted as 
subcritical by the authors. For $\Ek=10^{-4}$, the authors identify 
only supercritical bifurcations (because they find ${\bf B}={\bf 0}$ to be
unstable). They however did not perform a systematic study to
characterise the dynamo bifurcation, nor determined how it is controlled by the
governing parameters. This is the object of the work reported here.

We present here a review of the results obtained and first
described in french in Morin 2005~[\refcite{Morin2005}]~(available online).
These results were subsequently presented at several international
conferences~[\refcite{AGUAPS}]. They were not published in english so far,
but have probably inspired more recent articles~[\refcite{PontySimitev}]. 
We should finally stress that while our results were originally obtained using
only the numerical code call ``Magic'' (developed by G.~Glatzmaier,
U.~Christensen and J.~Wicht), we have since then reproduced these results 
using the local ``PaRoDy''code. Simulations with this second
code have confirmed the results presented here.

\section{Numerical investigations}

In all the simulations presented here, we solve the MHD equations in the
Boussinesq approximation in a rotating 
frame between two concentric spheres ($r_o/r_i=0.35$). The
reference frame is set such that the velocity vanishes
on both spheres (no-slip boundaries), a temperature difference is
maintained across the shell, and both the inner and the outer
domain are assumed to be electrically insulating.
The equations governing the solenoidal velocity $\bfu$ and magnetic $\bfB$
fields and the temperature $T$ can be written in non-dimensional form 
\begin{eqnarray}
{\rm E} \, 
\left[\partial _t \bfu +  (\bfu \cdot \bfnabla) \bfu \right]
&=& 
- \bfnabla \pi 
+ {\rm E} \, \Delta \bfu
- 2 \bfe_z \times \bfu 
\nonumber\\
&+& {\rm Ra} \, T \, \bfr 
+ {\rm Pm}^{-1} \left(\bfnabla \times \bfB \right) \times \bfB\, , \,\,\,\,\,\,
\end{eqnarray}   
\ \vskip -11mm
\begin{equation}
\partial _t \bfB = \bfnabla \times (\bfu \times \bfB) 
+ {\rm Pm}^{-1} \, \Delta \bfB \, ,
\label{induction}
\end{equation} 
\ \vskip -11mm
\begin{equation}
\partial_t T + (\bfu \cdot \bfnabla) T
= {\rm Pr}^{-1} \Delta T\, ,
\end{equation} 
using the previously introduced
Ekman number ${\rm E}={\nu}/{\Omega L^2}\, ,$
the (modified) Rayleigh number ${\rm Ra}={\alpha g \Delta T L}/{\nu \Omega}\, , $
the Prandtl number ${\rm Pr}={\nu}/{\kappa}\, , $ and
the magnetic Prandtl number ${\rm Pm}={\nu}/{\eta}\, ,$
with $L=r_o-r_i$, and $g=g_o/r_o$ where $g_o$ is the gravity at $r=r_o$.
For simplicity, we use here the same non-dimensional form as in~[\refcite{Christensen}].
The rapid rotation of the reference frame is measured by the smallness of $\Ek$.
In the following, the Prandtl number is set to unity, and thus ${\rm q}\equiv \kappa / \eta
= \Pm$.

As we will present our results using $\Ray/\Ray_c$ as controlling parameter, where
$\Ray_c$ is the critical Rayleigh number for the onset of thermal
convection, we want to recall, for completeness, that the values of $\Ray_c$
are respectively $55.9$, $60.8$, and $69.7$ for 
the Ekman numbers considered here i.e. $10^{-3}$, $3\cdot10^{-4}$, and
$10^{-4}$~.

\subsection{Influence of the magnetic Prandtl number on the dynamo bifurcation}    

\begin{figure}
\centerline{
a.\includegraphics[width=0.45\linewidth,clip=true]{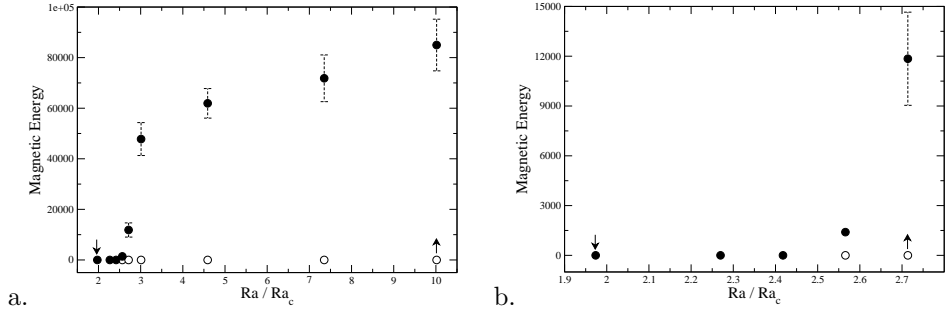}
\hfill
b.\includegraphics[width=0.45\linewidth,clip=true]{FIG/g_E3e-4q6_bif_zoom.eps}
}
\caption{Mean magnetic energy as a function of the Rayleigh number. The
  bifurcation is found to be supercritical for $\Ek=3.10^{-4}$ and
  $\Pm=6$.  Error bars indicate the standard deviation of energy
  fluctuations around the mean value. Graph b is an enlarged
  version of graph a, centered on the dynamo bifurcation.}
\label{E3e-4sup}
\end{figure}

Let us first fix the Ekman number and investigate the nature of the dynamo
bifurcation varying the magnetic Prandtl number.
With an Ekman number of $\Ek=3.10^{-4}$, when the magnetic Prandtl number is set to 
$\Pm=6$, the bifurcation diagram obtained numerically is reported on
figure~\ref{E3e-4sup}. 
We use filled dots to indicate stable solutions, and opened dots to
indicate unstable solutions. The diagram reports magnetic energies, for
varying values of the normalised Rayleigh number, dots indicate the time averaged values of the magnetic
energy, while the standard deviation is indicated using error bars.
The simulations are initialised with a small temperature perturbation,
a vanishing velocity, and an infinitesimal magnetic perturbation.

One can note on figure~\ref{E3e-4sup} that the first point with non-zero
magnetic field is obtained for $\Ray=2.57\times \Ray_c$, the previous point on
the diagram $\Ray=2.42\times \Ray_c$, corresponds to a stable ${\bf B}={\bf 0}$ state.
Figure~\ref{E3e-4sup}.a demonstrates how the time averaged magnetic energy
continuously tends to zero as he controlling parameter is decreased.
This demonstrates the supercritical nature of this bifurcation.

The onset of dynamo action is here obtained close to the onset of
convection. At the onset, the dynamo reaches a regime with constant kinetic
and magnetic energy.

We should stress that each point on such diagram, corresponds to
simulations of a few 100 of hours on a parallel computer. Computations are
three-dimensional and need
to be carried over a long enough time to obtain well converged averaged values.

\begin{figure}
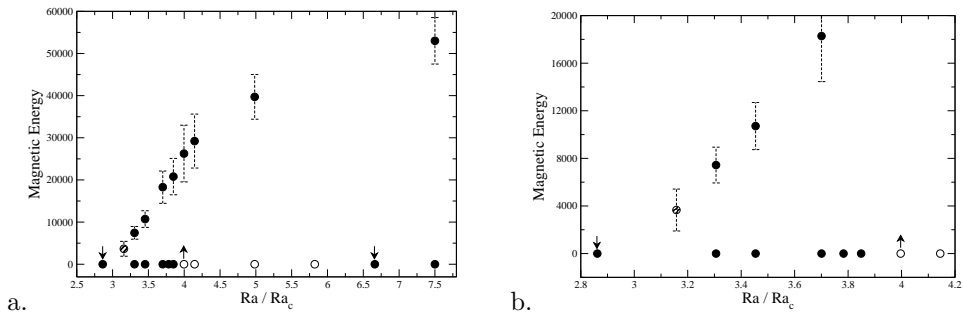

\begin{center}
a.\includegraphics[width=0.45\linewidth,clip=true]{FIG/g_bifdyn_q3_E3e-4.eps}
\hfill 
b.\includegraphics[width=0.45\linewidth,clip=true]{FIG/g_bifdyn_q3_E3e-4_zoom.eps}
\caption{Mean magnetic energy as a function of the Rayleigh number. This subcritical
  bifurcation was obtained for $\Ek=3.10^{-4}$ and $\Pm=3$.  Error bars
  indicate the standard deviation of the magnetic energy fluctuations
  around its mean value (dots). A close-up on the dynamo threshold is presented
  in graph b.}
\label{E3e-4sous}
\end{center}
\end{figure}

When the magnetic Prandtl number is decreased to 
$\Pm=3$, the onset of dynamo action occurs between 
$\Ray=3.85\times \Ray_c$ and $\Ray=4\times \Ray_c$ (further remote from the onset
of convection). The magnetic energy reported on figure~\ref{E3e-4sous}
exhibits an hysteretic behaviour
characteristic of a subcritical bifurcation.
The lower point on the subcritical branch 
is ``metastable''. It corresponds to $\Ray=3.16\times \Ray_c$ and
is designated on figure~\ref{E3e-4sous} by a stripped dot. By
metastable, we mean a self-sustained dynamo solution, which could be maintained
for a long time compared to viscous and ohmic diffusion
times,  but which eventually
collapses 
when the system experiences a sufficiently large fluctuation (in the sense of
a deviation from the mean magnetic energy). Once the
dynamo solution is lost, the noise being multiplicative, the system remains
in this state. 
In the case of this example, the
dynamo solution was maintained for $40$ viscous times,
i.e.~approximately $17$ magnetic decay times before collapsing.

\begin{figure}
\begin{center}
\includegraphics[width=0.45\linewidth,clip=true]{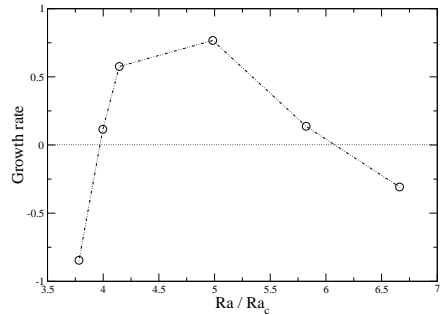}
\caption{Growthrate of the magnetic energy of a perturbation versus
  normalised Rayleigh number for $\Ek=3.10^{-4}$ and $\Pm=3$. The growthrate
  of the solution first increases as the value of the Rayleigh number is increased
  above the dynamo threshold. It reaches a maximum value, and then
  decreases to eventually become negative. The hydrodynamic solution
  restabilises above $\Ray\simeq 6\times \Ray_c$.} 
\label{E3e-4q3txcroiss}
\end{center}
\end{figure}

An intriguing characteristic of this bifurcation diagram is that the state
${\bf B}={\bf 0}$ restabilises for large enough values of the Rayleigh
number, above the dynamo threshold. 
This is not due to an isolated point which behaves unexpectedly. In fact
one can compute the linear growthrate of the magnetic field before
reaching saturation, and the decay rate when ${\bf B}={\bf 0}$ is stable
(see figure~\ref{E3e-4q3txcroiss}). 
This growthrate first increases above the onset, but then reaches a maximum as the Rayleigh
number increases and eventually becomes negative again. 

\begin{figure}
\centerline{
a.\includegraphics[width=0.45\linewidth,clip=true]{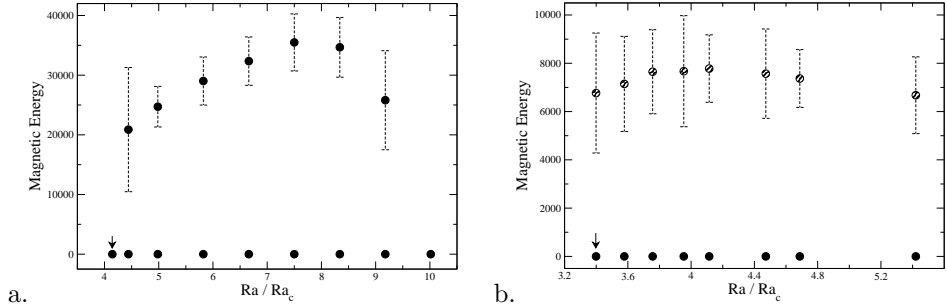}
\hfill
b.\includegraphics[width=0.45\linewidth,clip=true]{FIG/g_E3q3_Bif.eps}
}
\caption{Mean magnetic energy versus Rayleigh numbers. Isola
  diagram were obtained for $\Ek=3.10^{-4}$ and $\Pm=1.5$ (left plot) and
  $\Ek=10^{-3}$ and $\Pm=3$ (right plot).  Black dots (resp. stripped dots)
  indicate stable (resp. metastable) states. Error bars indicate the
  standard deviation of the magnetic energy fluctuations around its mean
  value. 
Metastable states are unstable to sufficiently large fluctuations
  of the velocity field. The probability for these fluctuations to occur
  are non zero. Purely hydrodynamic solutions were found to be always stable for this
  regime of parameters.}   
\label{ilots}
\end{figure}

A third type of bifurcation is obtained for a magnetic Prandtl number of 
$\Pm=1.5$ (see figure~\ref{ilots}.a). This bifurcation has an atypical
shape sometimes referred to as ``isola''. In this regime, the ${\bf B}={\bf
  0}$ solution does not destabilise, but a detached branch of dynamo
solution exists for $\Ray/\Ray_c \in [4.44,9.18]$

We have observed in this section, decreasing the magnetic Prandtl number from $\Pm=6$ to
$\Pm=1.5$, that we could describe supercritical, subcritical and isola
diagrams for dynamo action.
We shall note that the Rayleigh number had to be increased in this process 
(as the effects of Ohmic diffusion increase for decreasing values of the
magnetic Prandtl number).

\subsection{Influence of the Ekman number on the dynamo bifurcation}
\begin{figure}
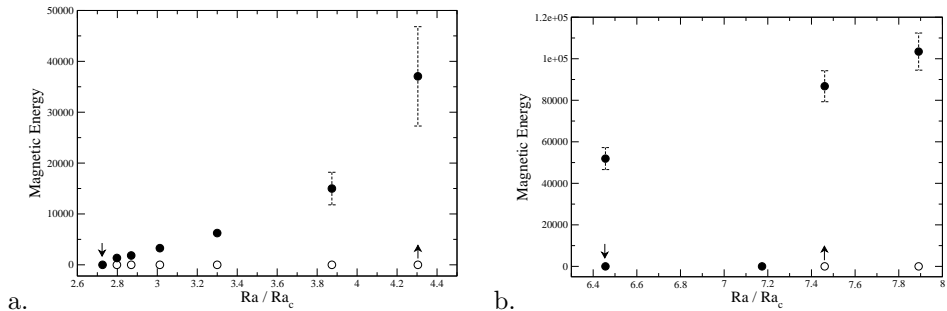

\centerline{
a.\includegraphics[width=0.45\linewidth,clip=true]{FIG/g_E4q3_bif.eps}
\hfill
b.\includegraphics[width=0.45\linewidth,clip=true]{FIG/g_E4q0p67_bif.eps}
}
\caption{Mean magnetic energy versus Rayleigh number. These supercritical
  (a) and subcritical (b) bifurcations were  obtained respectively for
  $\Ek=10^{-4}$ and $\Pm=3$ and $\Ek=10^{-4}$ and $\Pm=0.67$. Error bars
  indicate the standard deviation of the magnetic energy fluctuations
  around its mean value. }
\label{E4q3sup}
\end{figure}

Rather than studying the parameter space at fixed Ekman number and varying
the magnetic Prandtl number, we will now vary the Ekman number for a fixed magnetic Prandtl
number $\Pm=3$. 
Results obtained for $\Ek=10^{-3}$ are reported on
figure~\ref{ilots}.b. This diagram corresponds again to an isola. All dynamo
solutions were found to be metastable. The extreme points on this branch
were maintained for roughly $6 \, \tau_\nu$ other points were maintained in
time for up to $140 \, \tau_\nu$.
We have seen previously that the dynamo bifurcation is subcritical for an
Ekman number $\Ek=3.10^{-4}$ and a magnetic Prandtl number $\Pm=3$.
Let us now turn to $\Ek=10^{-4}$, the resulting bifurcation diagram is
presented on figure~\ref{E4q3sup}.a and corresponds to a supercritical bifurcation.
 
By increasing the Ekman number at fixed magnetic Prandtl number, we
therefore successively described supercritical, subcritical and isola
bifurcation diagrams. 
We further extended our study to
$\Ek=10^{-4}$ and $\Pm=0.67$ (figure~\ref{E4q3sup}.b) and obtained a subcritical bifurcation.

\subsection{Interpretation}

\begin{figure}
\begin{center}
\includegraphics[width=0.65\linewidth,clip=true]{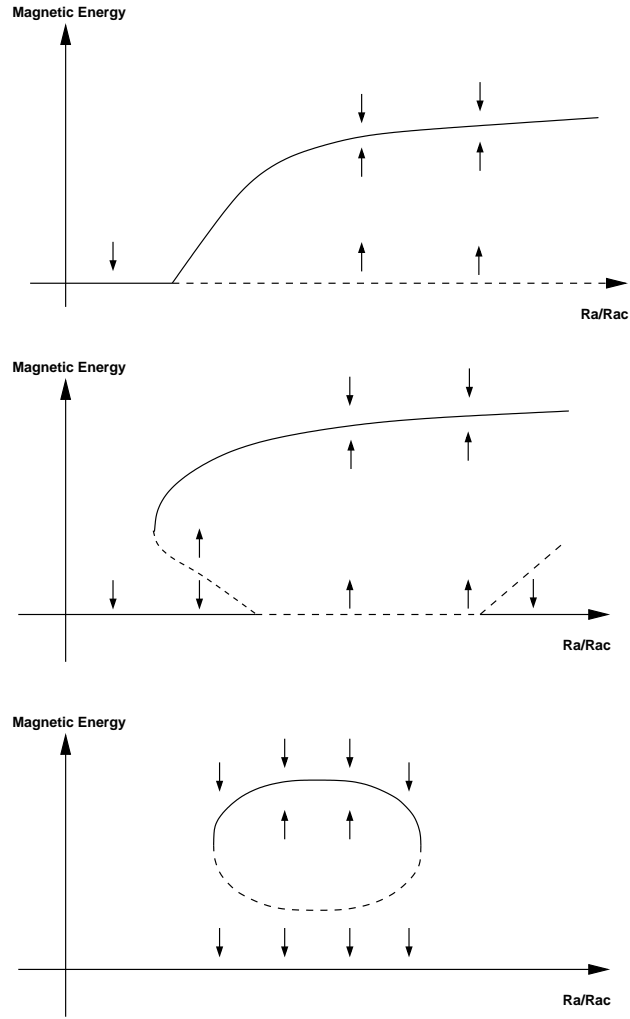}
\caption{Summary of the typical bifurcations obtained numerically. As the
  magnetic Prandtl number is decreased at fixed Ekman number one obtains
  successively supercritical (top), subcritical (middle) with
  restabilisation, and isola (bottom). The same sequence is achieved by
  increasing the Ekman number at fixed Rossby number.
  The dashed lines indicating unstable branches are here speculative,
  except for the ${\bf B}={\bf 0}$ state.}
\label{bif_mag}
\end{center}
\end{figure}

We have studied different bifurcations obtained for Ekman number values ranging
from $10^{-3}$ to $10^{-4}$, magnetic Prandtl number values from $0.67$ to $6$ and
Rayleigh number values from $\Ray\simeq 2\times \Ray_c$ to $\Ray\simeq10\times
\Ray_c$.  In this parameter regime, the evolution of the bifurcation diagram
is summarised 
in figure~\ref{bif_mag}. For a given Ekman number, a supercritical
bifurcation (top plot) is obtained for a sufficiently high value of the
magnetic Prandtl number. By decreasing its value, it is possible to obtain a
subcritical bifurcation (middle plot), which may exhibit unusual features,
such as re-stabilization of the purely hydrodynamical state.
An unstable branch therefore must exist, it could be connected, for larger values of the Rayleigh
number, to the stable branch corresponding to dynamo solutions. If the
magnetic Prandtl number is further decreased, the range of Rayleigh number for which
the non dynamo solution is unstable vanishes. An isola is then obtained
(bottom plot), in this situation the purely hydrodynamical solution is
always stable. The very same sequence is obtained by increasing $\Ek$ at
fixed $\Pm$, as the dipolar domain (domain II on figure~1) shifts towards
lower values of $\Pm$ as $\Ek$ is decreased.

\subsection{Feedback and coupling}
\begin{figure}
\centerline{
\includegraphics[width=0.45\linewidth,clip=true]{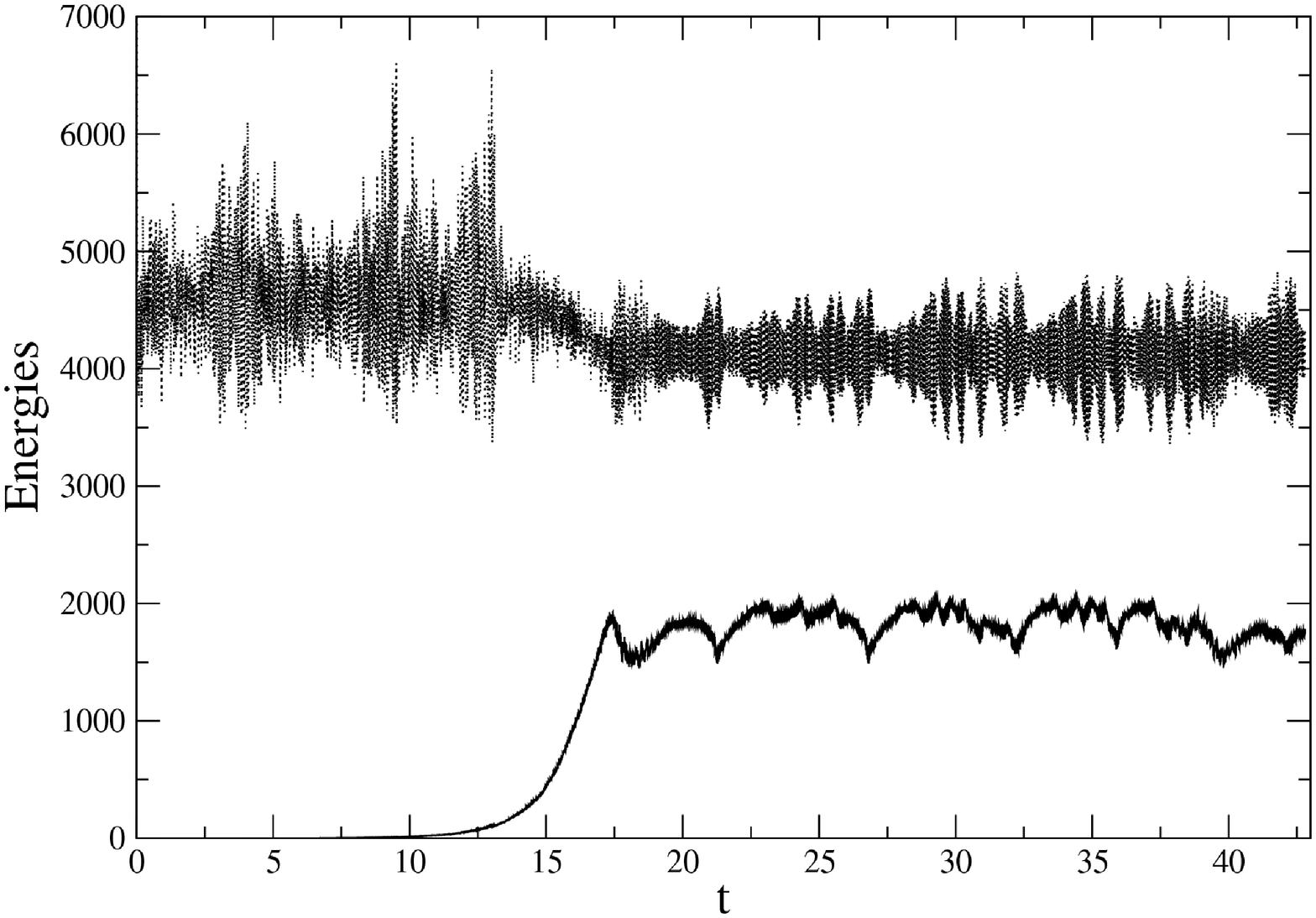}
\includegraphics[width=0.45\linewidth,clip=true]{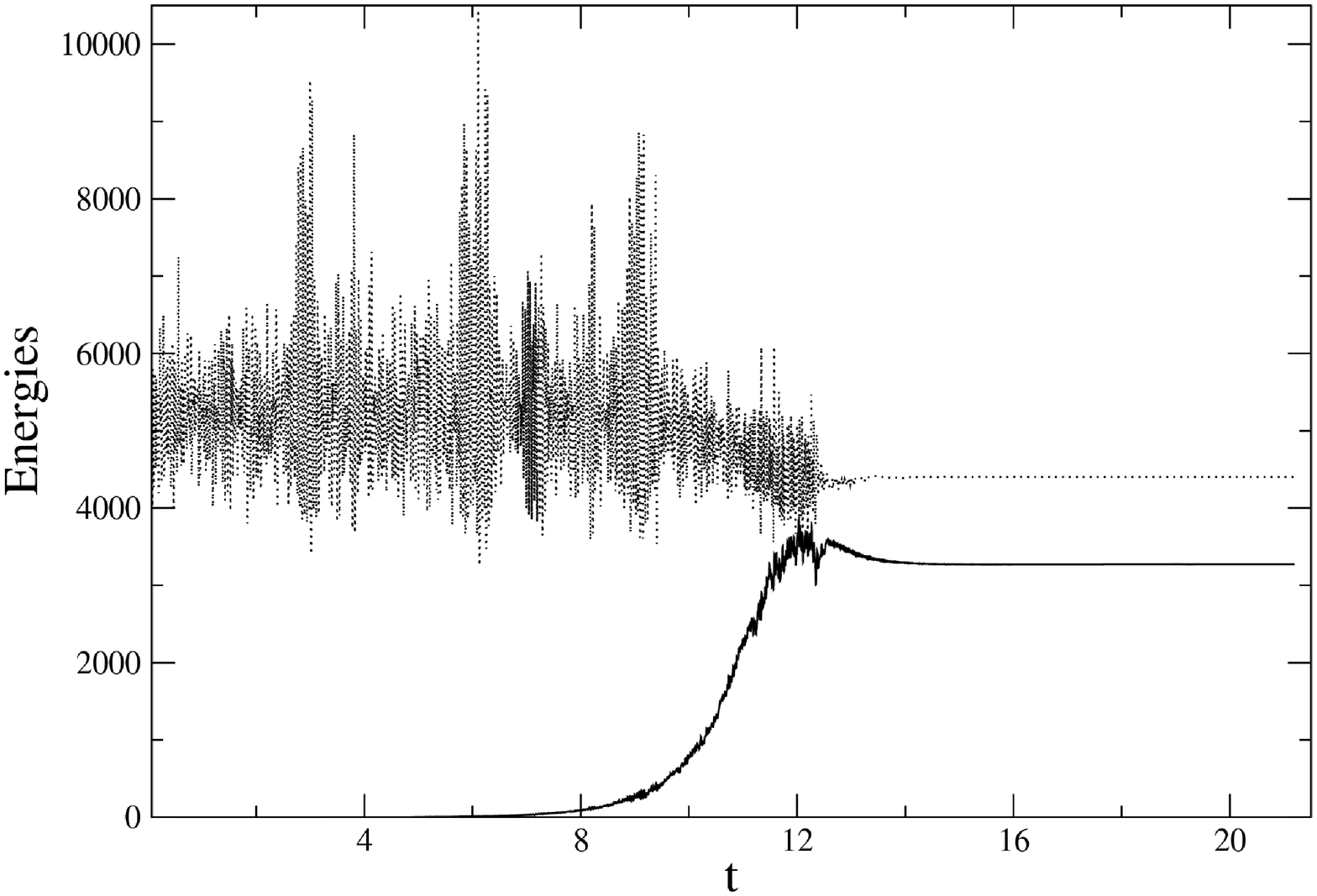}}
\centerline{
\includegraphics[width=0.45\linewidth,clip=true]{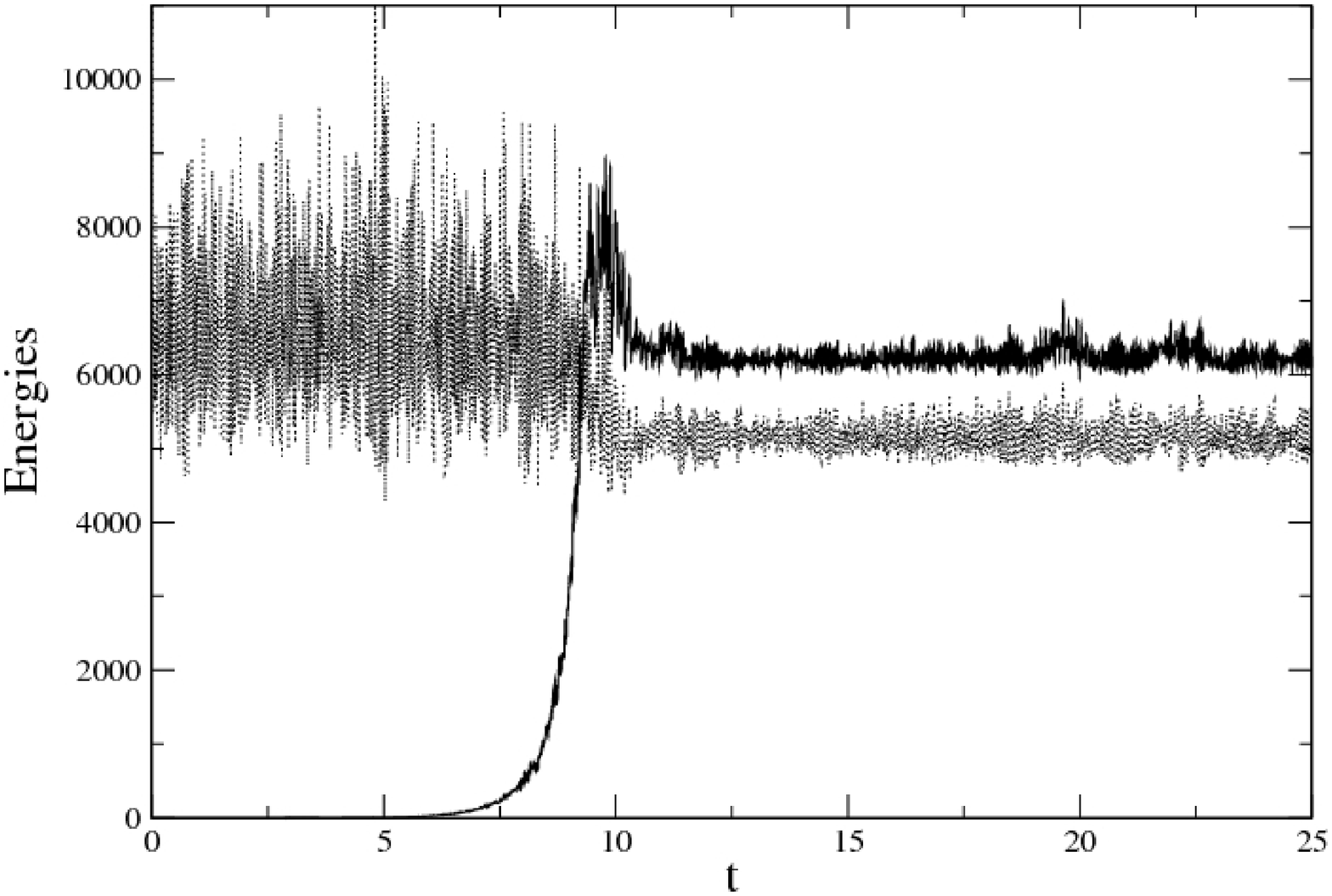}
\includegraphics[width=0.45\linewidth,clip=true]{FIG/g_E4_Ra300.eps}}
\caption{Magnetic energy (solid lines) and kinetic (dashed lines) as a function
  of time for $\Ek=1.10^{-4}$, $\Pm=3$ and from left to right and top to bottom
  $\Ray=2.87, 3.01, 3.30$ and $4.30\times \Ray_c$.}
\label{E4supBV}
\end{figure}

Let us now turn to the effects of the saturated magnetic field on the flow.
We compare on figure~\ref{E4supBV} simulations obtained for $\Ek=1.10^{-4}$ and
$\Pm=3$, before and after saturation. The magnetic field clearly acts here to
inhibit thermal convection. Not only does it lower the averaged value of
the kinetic energy, but it also reduces the amplitude of fluctuations.
This behavior is typical of the parameter space we investigated, there are
however a few noticeable exceptions.
 
\begin{figure}
\begin{center}
\includegraphics[width=0.45\linewidth]{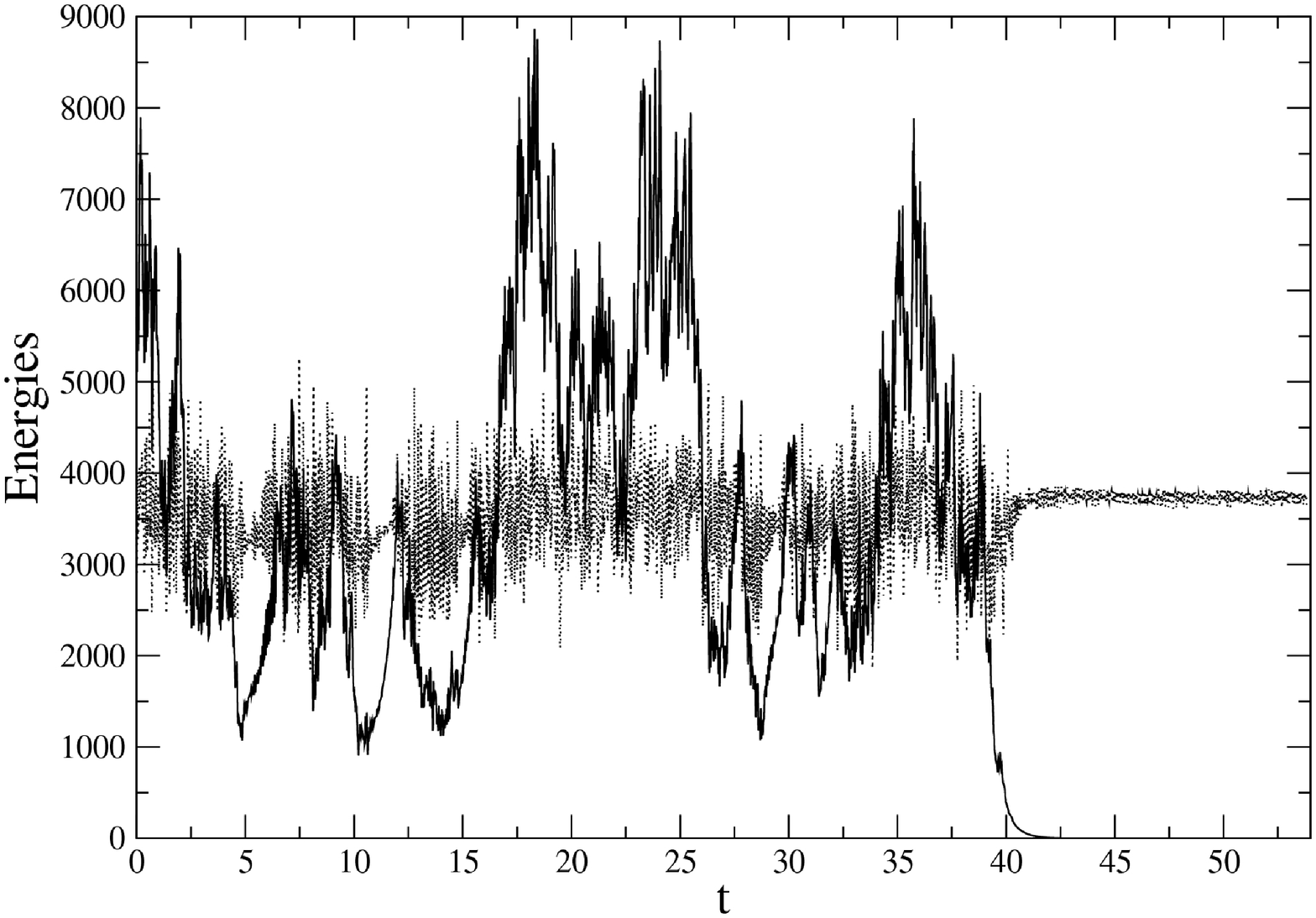}
\includegraphics[width=0.45\linewidth]{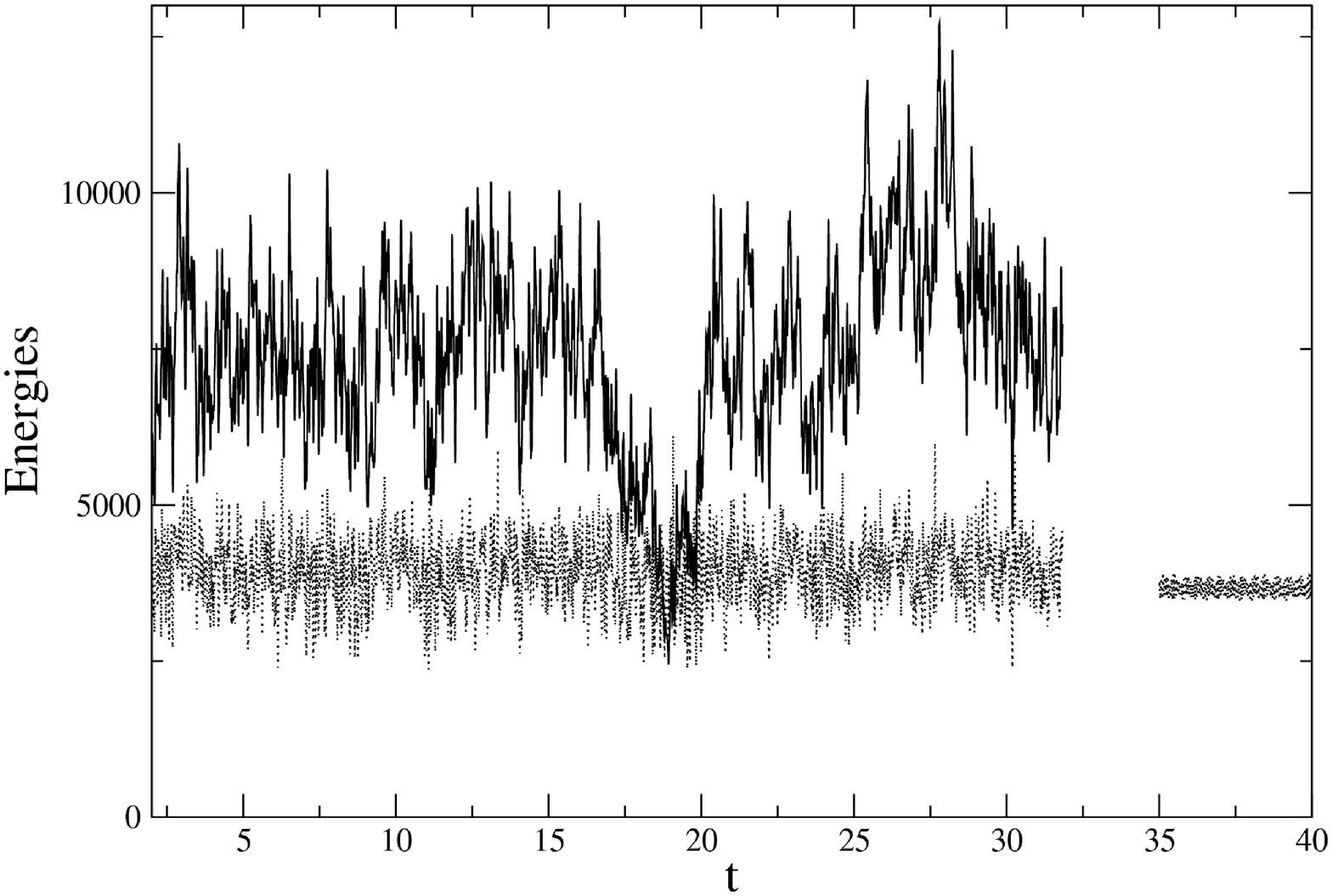} \\
\includegraphics[width=0.45\linewidth,clip=true]{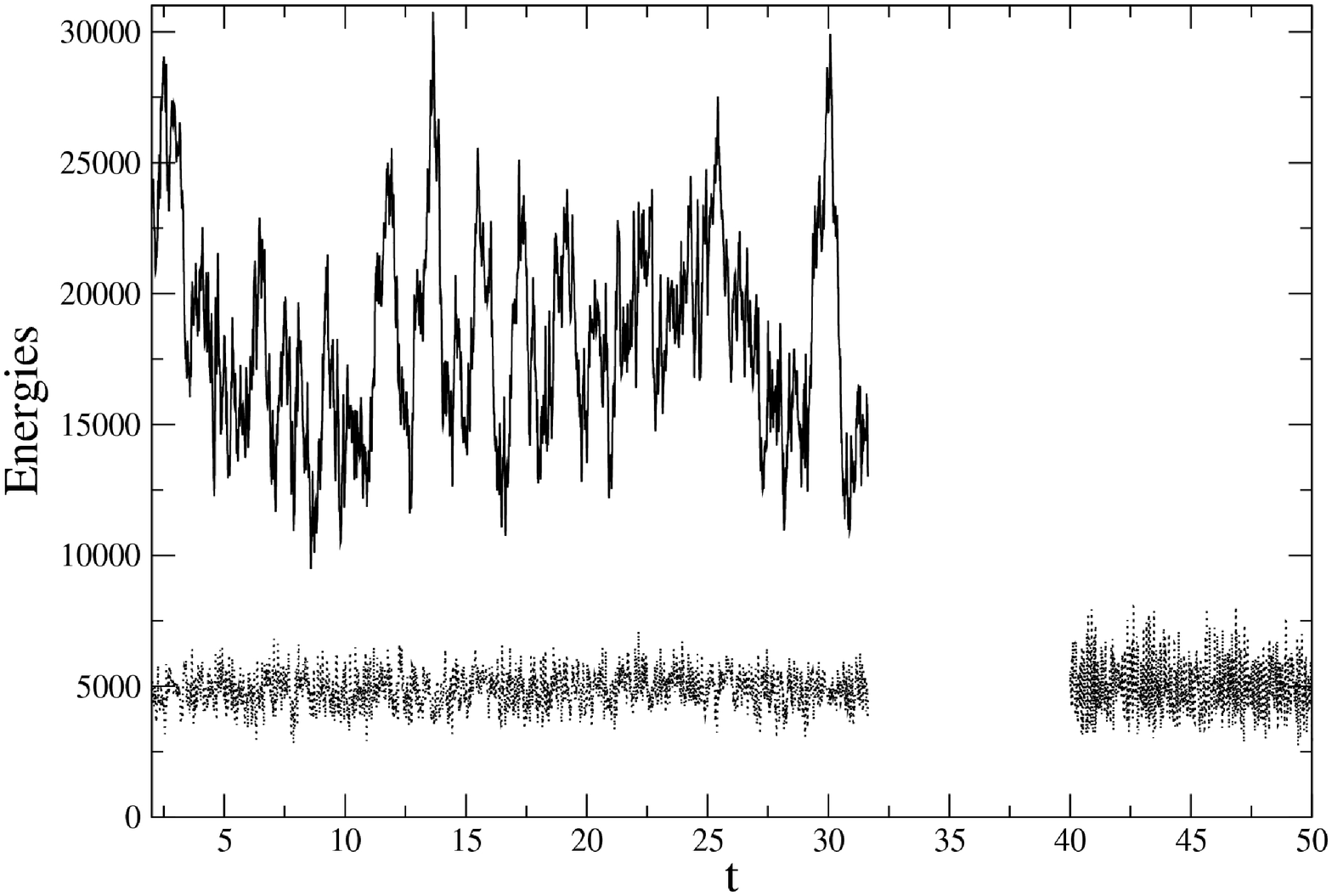}
\includegraphics[width=0.45\linewidth,clip=true]{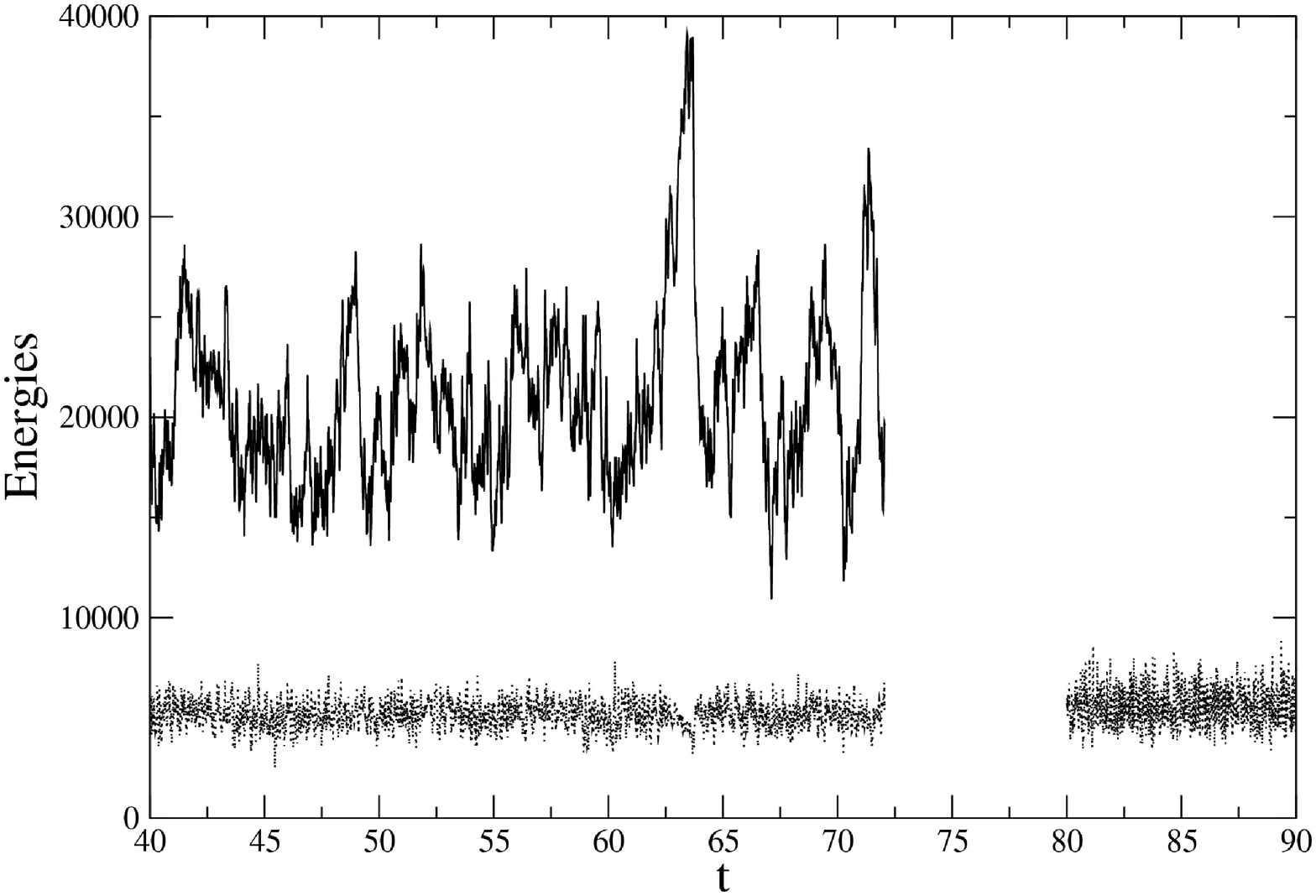}
\caption{Magnetic (solid) and kinetic (dashed) energies as a function of time
for $\Ek=3.10^{-4}$ and $\Pm=3$ and from top to bottom and left to right
$\Ray=3.16, 3.30, 3.70$ et $3.85\times \Ray_c$. For all graphs, the signal on
the far right part of the graph corresponds to the kinetic energy evolution
in the absence of magnetic field (purely hydrodynamic solution).}
\label{E3e-4q3sousBV}
\end{center}
\end{figure}

In the case of the subcritical bifurcation obtained for 
$\Ek=3.10^{-4}$ and $\Pm=3$, we can investigate a range of Rayleigh numbers
from $\Ray=3.16$ to $3.30 \times \Ray_c$ (which are lower than the linear
threshold value). It is then found, figure~\ref{E3e-4q3sousBV}, that the 
presence of the magnetic field yields stronger velocity fluctuations. 
Indeed the purely hydrodynamical solution (without magnetic field) is
presented on the right of each graph (either after a loss of dynamo action
in the metastable case $\Ray=3.16\times \Ray_c$, or through an independent
non-magnetic simulation).

\begin{figure}
\begin{center}
\includegraphics[width=0.45\linewidth,clip=true]{FIG/g_E3q3_Ra210_emag.eps}
\includegraphics[width=0.44\linewidth,clip=true]{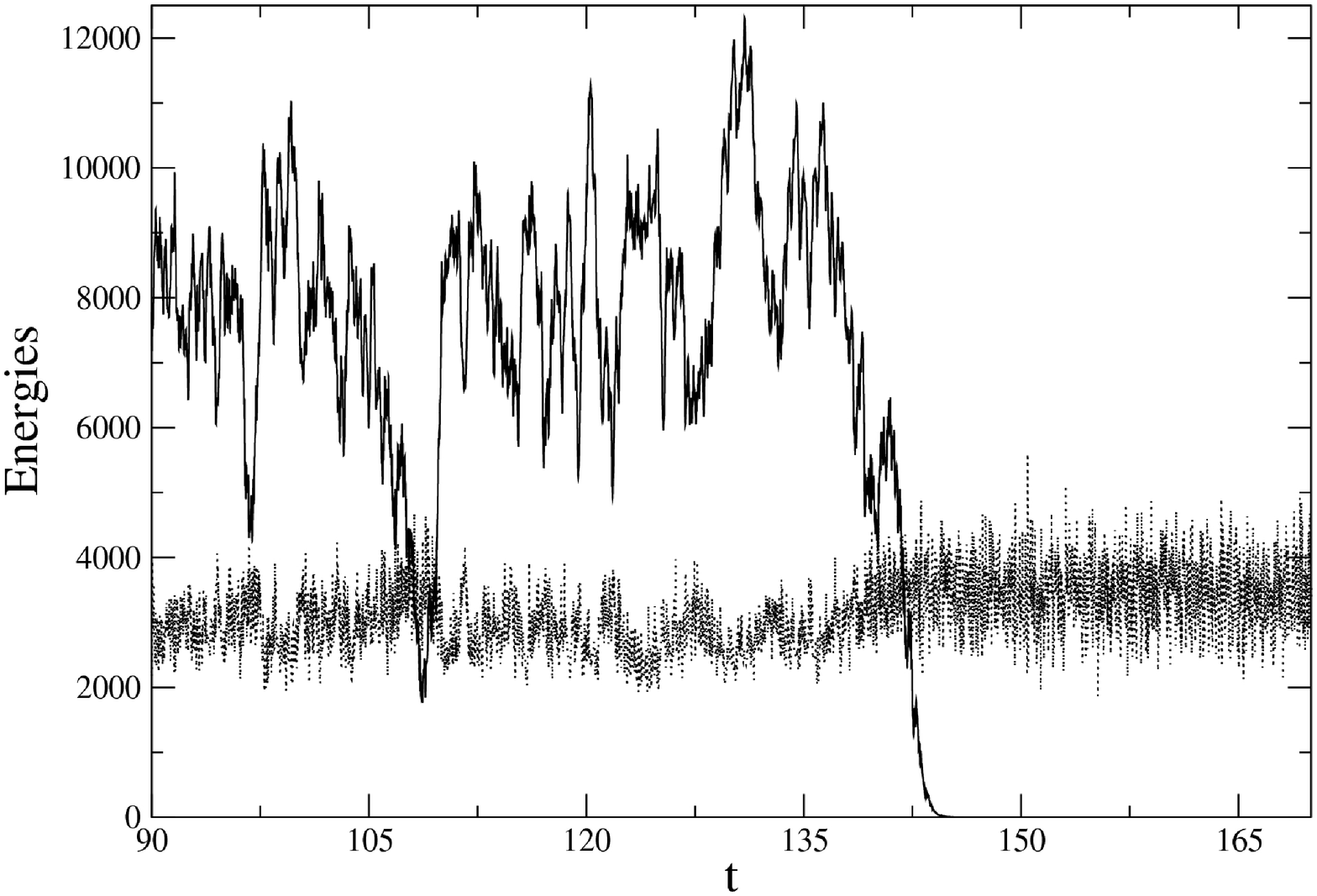}
\caption{Magnetic (and kinetic on the right graph in dashed lines) energies
  versus time for $\Ek=10^{-3}$, $\Pm=3$ and $\Ray=210$.
  The dynamo solution is lost after some $140\,\tau_\nu$.
  Both the mean and the fluctuating kinetic energies increase after the
  magnetic field collapse.}
\label{long}
\end{center}
\end{figure}

Dynamo solutions corresponding to the isola branch described for $\Ek=10^{-3}$
and $\Pm=3$ are metastable. Magnetic energy fluctuations are here comparable
to the mean value, which leads to a loss of dynamo action after a time
which depends on the Rayleigh number as well as on the particular choice of
initial conditions. We should highlight a simulation performed with a 
Rayleigh number $\Ray=3.76\times \Ray_c$, presented on figure~\ref{long}.
Dynamo action was here maintained for $140 \, \tau_\nu$ (with
$\tau_\nu=L^2/\nu$) before being lost, this corresponds to $20 \, \tau_\eta$ (with
$\tau_\eta=r_o^2/\eta$), and to $\simeq 200$ dipole decay time
($\tau_d=\tau_\eta / \pi^2$).
Such behaviour raises questions as to ``how long should a dynamo be
investigated to assess its stability?''.
Once lost, the dynamo solution cannot be recovered, as the purely
hydrodynamic solution is stable for this parameter regime. Once in the
basin of attraction of the ${\bf B}={\bf 0}$ state, the field cannot
recover. This is due to the multiplicative character of hydrodynamic
fluctuations in the induction equation. The effect of the hydrodynamic fluctuations
on the field vanishes near ${\bf B}={\bf 0}$.

Let us now turn to the statistical properties of the fluctuations of
kinetic and magnetic energies.
Figure~\ref{q} reports the probability density functions (PDF) of
both energies for $\Pm=3$ and on the top row
$\Ek=3.10^{-4}$, $\Ray=3.7\times \Ray_c$ (subcritical bifurcation), on the
bottom row $\Ek=10^{-4}$, $\Ray=3.3\times \Ray_c$ (supercritical bifurcation). 
In both cases the PDFs of the magnetic energy are much less symmetrical than their
counterpart on the kinetic energy. This is a robust characteristic, which
was observed in all our simulations.

This appears again to be a direct consequence of the multiplicative character of the
hydrodynamic fluctuations in the induction equation. 
Whereas temperature fluctuations act as additive noise on the velocity
equation, velocity fluctuations act as a multiplicative noise on the
magnetic field (i.e. a term proportional to $\bf B$).
One can show on a very simple system how a multiplicative 
white noise (with symmetric PDF) yields a chi-2 law
(asymmetric PDF) for the variable affected by the noise.
Let us consider the temporal evolution of a scalar, 
for a conservative system, we can write
\begin{equation}
\frac{\dd X}{\dd t}=  -\frac{\dd G}{\dd X} \, .
\label{pot2}
\end{equation} 
An additive noise would drive fluctuations of $X$ within a fixed potential
$G$, a multiplicative noise will instead modify the shape of the potential.
Let us consider, for example, the case of a steady subcritical bifurcation
in the presence of a multiplicative noise. The normal form (assuming the
$X\mapsto -X$ invariance) yields a potential of the form
\begin{equation}
G(X)=-\frac{\mu + \xi(t)}{2}\,X^2 - \frac{\alpha}{4}\,X^4+\frac{1}{6}\,X^6 \, .
\label{le_potentiel}
\end{equation}
In the absence of noise ($\xi\equiv 0$), solutions to 
(\ref{pot2}-\ref{le_potentiel}) are: the trivial $X=0$ (stable for $\mu<0$ unstable
otherwise), and the four roots of
\mbox{$\mu+\alpha X^2-X^4$,} for
$-\alpha^2/4\leqslant \mu<0$. Two of these are unstable, the two others
being stable (subcritical branch) and continuous for $\mu\geqslant0$~.    

\begin{figure}
\begin{center}
\includegraphics[width=0.5\linewidth,clip=true]{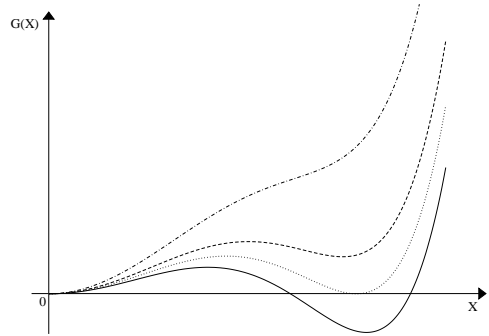}
\caption{Potential $G(X)$ as a function of the order parameter $X$. The
potential represented here takes the form (\ref{le_potentiel}) setting
  $\alpha=5$ and the prefactor of the $X^2$ term to
  $2$ (solid line), $2.3435$ (dotted), $2.7$ (dashed) et $3.7$ (dot-dashed).} 
\label{potbruit}
\end{center}
\end{figure}

\begin{figure}
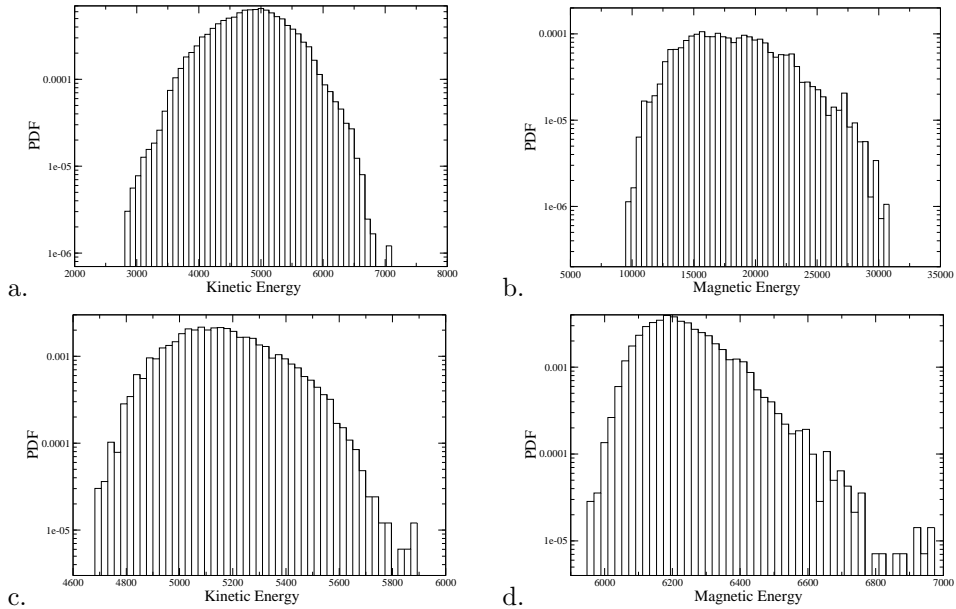

\centerline{
a.\includegraphics[width=0.45\linewidth,clip=true]{FIG/g_E3e-4q3Ra225_PDF_Ekin.eps}
\hfill
b.\includegraphics[width=0.45\linewidth,clip=true]{FIG/g_E3e-4q3Ra225_PDF_Emag.eps}}
\vskip 2mm
\centerline{
c.\includegraphics[width=0.45\linewidth,clip=true]{FIG/g_E4q3Ra230_PDF_Ekin.eps}
\hfill
d.\includegraphics[width=0.45\linewidth,clip=true]{FIG/g_E4q3Ra230_PDF_Emag.eps}}
\caption{Probability Density Functions for the kinetic energy (left) and
  magnetic energy (right), for  $\Pm=3$ and $\Ek=3.10^{-4}$, $\Ray=3.7\times \Ray_c$
  (top), $\Ek=10^{-4}$, $\Ray=3.3\times \Ray_c$ (bottom). The top line corresponds to a
  subcritical bifurcation, the bottom one to a supercritical bifurcation.}
\label{q}
\end{figure}

Let us now consider the subcritical branch of this 
system with noise. Below the threshold, the parameter $\mu$ is negative.
Figure~\ref{potbruit} examplifies the shape of the potential $G(X)$ for
different values of $\xi(t)$, and thus of the prefactor of the 
$X^2$ term. 
When it is low enough (solid line), there exists a negative minimum. This
minimum provides the preferred state of the system. When $\xi(t)$ becomes
large enough, the potential can fluctuate until both minima of the
potential reach the same value (dotted curve).
If the prefactor of the $X^2$ term becomes even larger, the potential
minimum associated to non-zero $X$ becomes local and the prefered state
of the system becomes $X=0$. 

It is important to appreciate that $X=0$ is an absorbing state. Indeed the noise being
multiplicative will not affect the system any longer when $X=0$.
If fluctuations are large enough, non-trivial solutions of a subcritical
system in the presence of multiplicative noise can thus be destabilised
through a ``large'' fluctuation. The trivial solution
then becomes permanent. This simpler model explains the behaviour reported above in our dynamo
simulations. 

The presence of a multiplicative noise implies that previously well
defined equilibrium values (in the absence of noise) become probabilities.
Let us consider a simple supercritical example of normal form
\begin{equation}
\frac{\dd X}{\dd t}= \left[\mu +\xi(t)\right]\,X- X^3 \, .
\label{surbruit}
\end{equation} 
If one assumes that $\xi(t)$ is a white and gaussian noise of variance $D$,
one can then show that the equation governing the evolution of the probability
$P$ of $X$ to have a given value is
\begin{equation}
\frac{\partial P}{\partial t} = -\frac{\partial}{\partial X}\left[(\mu X -
X^3)P\right] + \frac{D}{2} \frac{\partial}{\partial X}\left( X
\frac{\partial }{\partial X} XP \right) \, .
\end{equation}
Integrating this equation in $X$ for steady probabilities, yields
\begin{equation}
J = (-\mu X + X^3)\, P + \frac{D}{2} \left( X \frac{\partial }{\partial X} XP
\right)\, .
\end{equation}
The right-hand-side vanishes for $X=0$, which implies that
$J=0$. The change of variable $q=XP$ provides
\begin{equation}
\frac{\partial q}{\partial X} = \frac{2}{D} \left( \frac{\mu}{X} - X
\right) q \, ,
\quad \mbox{and thus}\quad
P = C\, |X|^{2(\mu/D) -1}\, e^{- X^2/D}\, . 
\end{equation}
It is worth noting that starting from a multiplicative gaussian white noise
(thus with a symmetric probability distribution), one obtains a non-symmetric
probability distribution for the variable $X$.

One can show in a similar manner that for a subcritical bifurcation with a
normal form given by (\ref{pot2}-\ref{le_potentiel}), the probability $P(X)$ to meet a
given value is
\begin{equation}
P = C\, |X|^{2(\mu/D) -1}\, e^{X^2/D- X^4/2D}\, . 
\end{equation}

This simple example can, off course, not account for the full complexity of
the coupled non-linear MHD system, yet it demonstrates how the structure of
the induction equation accounts for the tendency for a
stronger asymmetry in PDFs of the magnetic than in the kinetic energy (see figure~12).

\subsection{Multiple Solutions}
\begin{figure}
\begin{center}
\includegraphics[width=0.48\linewidth,clip=true]{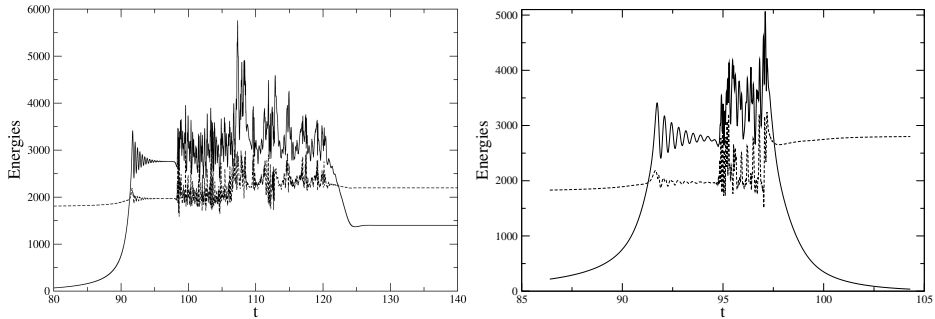}
\includegraphics[width=0.48\linewidth,clip=true]{FIG/g_E3e-4q6Ra156_real2.eps}
\caption{Magnetic (solid) and kinetic (dashed) energies versus time for
$\Ek=3.10^{-4}$, $\Pm=6$ and $\Ray=2.57 \times \Ray_c$. Left and right graphs
  correspond to two independent simulations with initial
  conditions which differ only from a very small perturbation (identified
  in the text as ``run 1'' and ``run 2'').}
\label{E3e-4q6Ra156Etot}
\end{center}
\end{figure}

In the course of the investigation reported here, we did not encounter proper multiple
co-existing branches of dynamo action (but see section~3). We however
observed an interesting behaviour which happens 
to be directly related to the non-uniqueness of the purely hydrodynamic
solution itself.
Both the magnetic and the kinetic energies are presented for
$\Ek=3.10^{-4}$, $\Pm=6$ and $\Ray=2.57 \times
\Ray_c$ in figure~\ref{E3e-4q6Ra156Etot}.
Two different simulations, which only differ by a minute initial fluctuation,
are presented on this figure. The behaviour is rather unexpected and untypical.
While the simulations are first similar, the dynamo solution appears to
reach a stable state (damped oscillations) but then violently
destabilises. In one case dynamo action is maintained (left), but it is lost in the
other one (right). 

\begin{figure}
\begin{center}
\includegraphics[width=0.495\linewidth,clip=true]{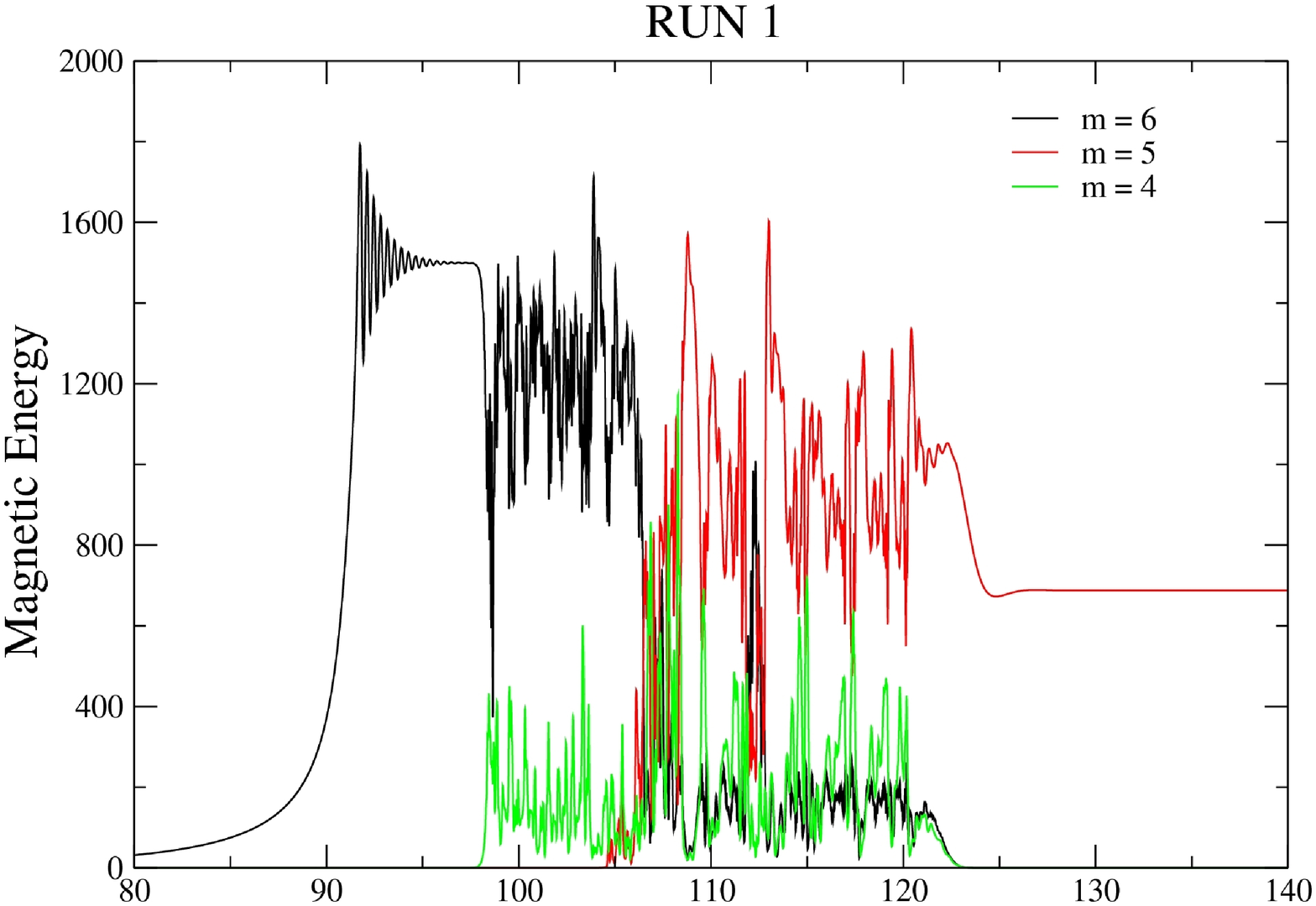}
\includegraphics[width=0.478\linewidth,clip=true]{FIG/g_E3e-4q6Ra156_B_mod_75.eps} \\
\includegraphics[width=0.495\linewidth,clip=true]{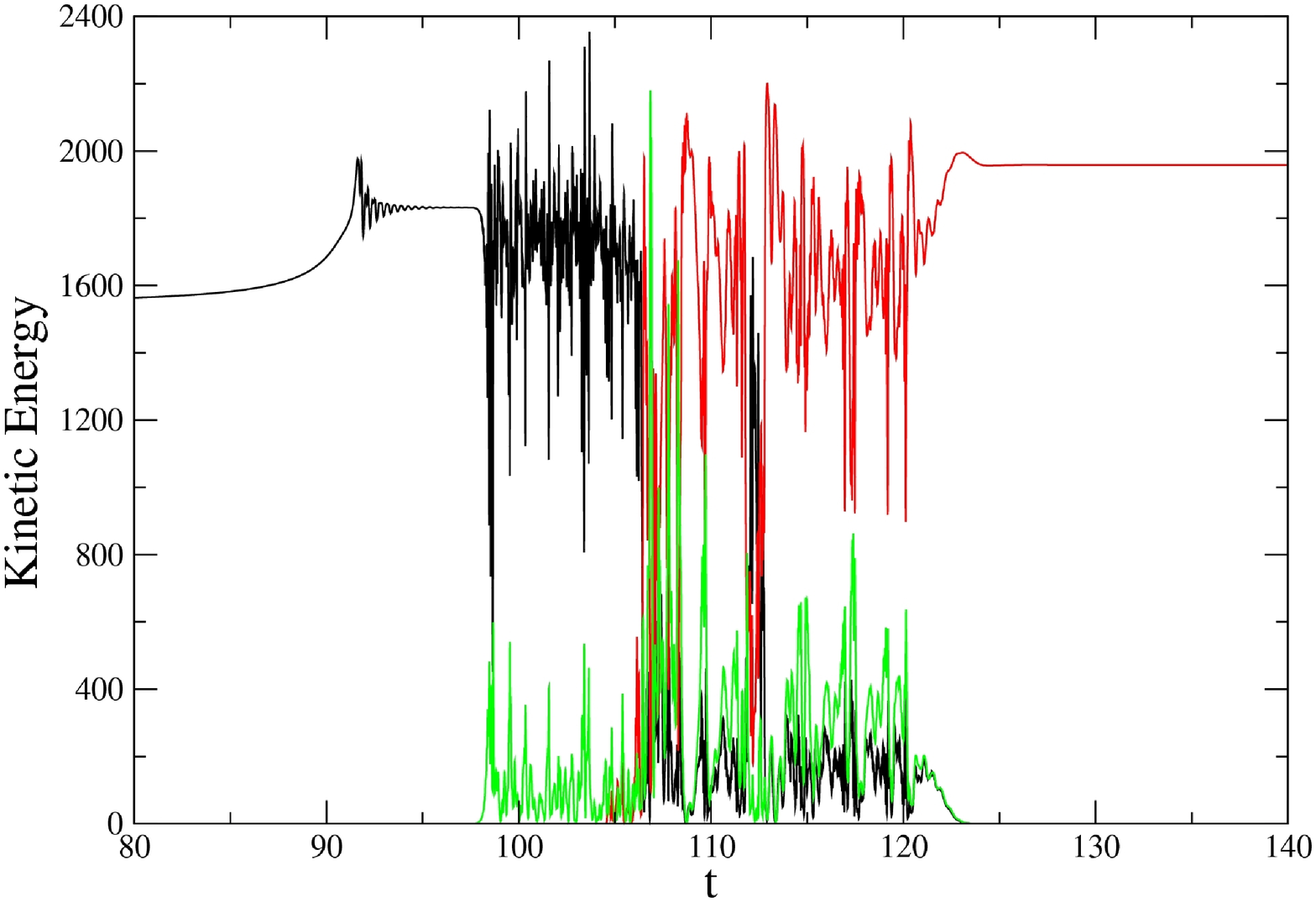}
\includegraphics[width=0.478\linewidth,clip=true]{FIG/g_E3e-4q6Ra156_V_mod_75.eps}
\caption{Magnetic (top) and kinetic (bottom) energies corresponding to the
  azimuthal Fourier modes $m=4,\, 5$ and $6$ for the simulations presented
  on figure~\ref{E3e-4q6Ra156Etot}.}
\label{E3e-4q6Ra156_2D}
\end{center}
\end{figure}

For this parameter space, we identified at least two stable solutions to the purely hydrodynamic problem.
Further more, we observed (figure~\ref{E3e-4q6Ra156_2D}),
 a modification in time of the dominant azimuthal Fourier mode
both for the kinetic and magnetic fields. Simulations were initialised
with a random perturbation of the magnetic and temperature fields and a
fluid at rest. The chosen parameters are here extremely close to the
threshold value, and the magnetic field growth is slow (the growthrate is
$0.058$). Thermal convection therefore had enough time to develop largely
uninfluenced by the Lorentz force. For this parameter regime the kinetic
energy of the most unstable mode of hydrodynamic convection is steady in
time and the 
dominant modes are $m=6$ and $m=0$.
When the field reaches saturation is has the same structure as the flow,
dominated by an $m=6$ symmetry.
This mode however quickly destabilises and, as illustrated on
figure~\ref{E3e-4q6Ra156_2D}, after a competition between the 
$m=4,\,5$ and $6$ modes, the dominant modes becomes (both for the kinetic and
the magnetic energy) $m=5$ for the first run and $m=4$ for the second one.
The simulation ``run 2'', corresponds to a loss of dynamo
action, as the $m=4$ convection mode does not appear able to sustain the
field against diffusion. 
Both simulations (``run1'' and ``run 2'') were carried further in time for
another $30 \, \tau_\nu$, to assess the stability of their final state.

This intricate behaviour therefore comes as the consequence of the
existence of three possible
modes of hydrodynamic convection (all stables). The fastest growing mode,
$m=6$ does act as a kinematic dynamo, but this dynamo is non-linearly
unstable and destabilises either to the $m=4$
or to the $m=5$-mode. The later is also a dynamo, but contrary to
$m=6$ it is a stable solution. The $m=4$-mode does not act as dynamo, the
magnetic field is then lost. 
We did not observe any heteroclynic cycle. The behavior reported here is
reminiscent of self-killing dynamos as described in~[\refcite{Fuchs}].

\section{Further developments}

We presented here a study of the onset of dynamo action over a wide
parameter range. We should note however that one parameter was, for
simplicity, kept constant throughout the entire study: the Prandtl number
(${\rm Pr}=\nu/\kappa$). It would be interesting to know how this parameter
affects the picture presented here.

Since this work was performed and published in french in
2005~[\refcite{Morin2005}], we have further investigated the parameter
space towards large values of the
magnetic Prandtl number. Yet another kind of bifurcation diagram has been
produced. It is characterized by two co-existing branches of different
energies.
These results will be presented and discussed elsewhere~[\refcite{Soumis}].

\section*{Acknowledgements}

The authors are very grateful to Stephan Fauve for many fruitful discussions
in the course of this work.


\section*{References}


\begin{thebibliography}{0}

\bibitem{KC02} C. Kutzner \& U. Christensen, From stable dipolar towards
  reversing numerical dynamos, Phys. Earth Planet. Int., {\bf 131}, 29,
  2002.  

\bibitem{Aubert} U. Christensen \& J. Aubert, Scaling properties of
  convection-driven dynamos in rotating spherical shells and application to
  planetary magnetic fields, Geophys. J. Int. {\bf 166},
  97-114, 2006.

\bibitem{CK03} C. Kutzner,  {\it Untersuchung von Feldumkehrungen an einem numerischen Modell
des Geodynamos}, PhD thesis, G\"ottingen University, 2003.

\bibitem{COG} U. Christensen, P. Olson \& G. Glatzmaier, Numerical modeling
  of the geodynamo: a systematic parameter study, Geophys. J. Int. {\bf 138},
  393, 1999.

\bibitem{CRAS} E. Dormy \& J.-L. Le Mou\"el, Geomagnetism and the
  dynamo: where do we stand?, C.R.Physique, {\bf 9}, 711-720, 2008.

\bibitem{Morin2005} V. Morin, Instabilit\'es et bifurcations associ\'ees \`a la
  mod\'elisation de la G\'eodynamo, PhD thesis, Universit\'e Paris VII
  (2005), (http://tel.archives-ouvertes.fr/tel-00011484/en/).

\bibitem{AGUAPS} 
Morin V., Dormy E., ``Numerical investigation of the dynamo bifurcation'',
FG.00008, APS, Division of Fluid Dynamics, 59th annual meeting, Tampa, USA, Nov. 2006;
Morin V., Dormy E., ``Dynamo bifurcation in geodynamo models'', \#DI31A-0255, AGU Fall
Meeting, San Francisco, USA, Dec 2007;
Dormy E., workshop ``Modelling geophysical systems by statistical mechanics
methods'', Erice, Italy, May 2008;
Morin V., Dormy E., ``Numerical investigation of the dynamo bifurcation in
geodynamo models'', EURO MHD, Nice, France, sept. 2008;
Dormy E., 
workshop ``Waves and instabilities in geophysical 
and astrophysical flows'',
Porquerolles, France, May 2009.

\bibitem{PontySimitev} Y. Ponty, J.P. Laval, B. Dubrulle, F. Daviaud,
  J.F. Pinton, PRL, {\bf 99}, 224501, 2007; R.D. Simitev, F.H. Busse, EPL,
  {\bf 85}, 19001, 2009. 

\bibitem{Christensen} Christensen {\it et al}, Phys. Plan. Earth Int. {\bf
  128}, 25-34, 2001.

\bibitem{Fuchs} H. Fuchs, K.H. R\"adler, M. Reinhardt, in Dynamo and
  Dynamics, Chossat, Armbuster \& Oprea Eds, Kluwer 2001.

\bibitem{Soumis} V. Morin \& E. Dormy, Weak and Stong Field Dynamos, 2009
  submitted. 
\end{thebibliography}
\end{document}